\documentclass[aps,prb,groupedaddress, twocolumn, notitlepage, portrait]{revtex4-2}
\usepackage[utf8]{inputenc}
\usepackage[T1]{fontenc}
\usepackage{amsmath}
\usepackage{amsfonts}
\usepackage{amssymb}
\usepackage{braket}
\usepackage{graphicx}
\usepackage[colorlinks=true, citecolor=blue, linkcolor=blue, urlcolor=blue]{hyperref}
\begin{document}
\title{Approximate Hofstadter- and Kapit-Mueller-like parent Hamiltonians for Laughlin states on fractals}
\author{B\l a\.{z}ej Jaworowski}
\affiliation{Department of Physics and Astronomy, Aarhus University, DK-8000 Aarhus C, Denmark}
\author{Michael Iversen}
\affiliation{Department of Physics and Astronomy, Aarhus University, DK-8000 Aarhus C, Denmark}
\author{Anne E. B Nielsen}
\affiliation{Department of Physics and Astronomy, Aarhus University, DK-8000 Aarhus C, Denmark}

\begin{abstract}
Recently, it was shown that fractional quantum Hall states can be defined on fractal lattices. Proposed exact parent Hamiltonians for these states are nonlocal and contain three-site terms. In this work, we look for simpler, approximate parent Hamiltonians for bosonic Laughlin states at half filling, which contain only onsite potentials and two-site hopping with the interaction generated implicitly by hardcore constraints (as in the Hofstadter and Kapit-Mueller models on periodic lattices). We use an ``inverse method'' to determine such Hamiltonians on finite-generation Sierpi\'{n}ski carpet and triangle lattices. The ground states of some of the resulting models display relatively high overlap with the model states if up to third neighbor hopping terms are considered, and by increasing the maximum hopping distance one can achieve nearly perfect overlaps. When the number of particles is reduced and additional potentials are introduced to trap quasiholes, the overlap with a model quasihole wavefunction is also high in some cases, especially for the nonlocal Hamiltonians. We also study how the small system size affects the braiding properties for the model quasihole wavefunctions and perform analogous computations for Hamiltonian models.
\end{abstract}
\maketitle

\section{Introduction}

In a classic work from 1977 \cite{leinaas1977theory}, Leinaas and Myrheim showed that, the topology of configuration space in one and two dimensions opens up a possibility for the existence of particles, which are neither bosons nor fermions. Later, such particles became known as ``anyons'' \cite{wilczek1982quantum}. Their full potential is revealed in two dimensions, where they can be exchanged without passing through each other. In particular, the non-Abelian anyons in two dimensions were proposed as a gateway to quantum computing \cite{kitaev2003fault,nayak2008nonabelian}. While no fundamental particle was found to obey anyonic statistics, anyons were observed experimentally as quasiparticle excitations of topological orders \cite{nakamura2020direct,bartolomei2020fractional,semeghini2021probing,satzinger2021realizing}. 

The integer dimensions do not exhaust all possible options. If we define the dimension of the system as the Hausdorff dimension, then non-integer dimensions can be found in fractal systems \cite{mandelbrot1983fractal}. Quantum systems with fractal geometry were already realized experimentally \cite{newkome2006nanoassembly,shang2015assembling,wang2019construction,kempkes2019design} and also simulated in photonic systems \cite{xu2021quantum,biesenthal2022fractal,biesenthal2022fractal}. 
Moreover, using optical tweezers, one can create atomic arrays of arbitrary shape \cite{barredo2016atom,barredo2018synthetic}, and proposals for implementing complex hopping in these systems are being developed \cite{weber2022experimentally,wu2022manipulating}. There is an ongoing effort to create arbitrary patterns of potentials and hoppings in quantum simulators \cite{aidelsburger2013realization, nogrette2014single,tai2017microscopy}, including pairwise-tunable long-range complex hoppings \cite{hung2016quantum}. The development of this field suggests that in the near future it may be possible to realize systems which are not only fractal-shaped, but also highly controllable and at the same time exhibit many-body effects. Thus, a question arises: can such systems host anyons? Some authors considered possible statistics on arbitrary graphs (which includes also graphs based on fractals) \cite{harrison2011quantum,harrison2014nparticle,maciazek2019nonabelian}. Another approach is to construct topologically ordered states on fractal lattices \cite{manna2020anyon,zhu2021topological,li2022laughlin}. In particular, it was shown that lattice equivalents of Laughlin fractional quantum Hall states host anyonic excitations when defined on fractals \cite{manna2020anyon}.

The model Laughlin states investigated in \cite{manna2020anyon} have an exact parent Hamiltonian that can be obtained from conformal field theory. While such a Hamiltonian certainly can be useful (we know its ground state for arbitrarily large systems and that this ground state is topologically ordered), its form is quite complicated as it consists of two- and three-site terms which are nonlocal (i.e.\ connect sites at arbitrary distance from each other). In contrast, quantum Hall states in periodic lattices have a simpler parent Hamiltonian: the Kapit-Mueller model \cite{kapit2010exact}, which contains only two-site, nonlocal hopping terms and the interaction. In the simplest case of filling $1/2$, the latter can be generated implicitly by the hardcore constraint. Even simpler is the Hofstadter model \cite{harper1955single,hofstadter1979energy} with hardcore interactions \cite{sorensen2005fractional}, where the hoppings are local (i.e.\ hopping from and to a given site is possible only in its vicinity). It is not an exact parent Hamiltonian for any nonzero flux, but at low flux its ground state has a high overlap with a model Laughlin state \cite{sorensen2005fractional}. This raises the question whether ground states of similar Hamiltonians on fractal lattices can exhibit topological orders in non-integer dimensions.

One can reformulate and narrow down this question by asking: for a given fractal lattice, is it possible to find a Hamiltonian similar to the Hofstadter or Kapit-Mueller model, whose ground state is either exactly given by the model Laughlin state \cite{manna2020anyon} or is approximated by it?  The problem of systematically finding parent Hamiltonians for a given target state has gained attention in recent years, and several numerical methods of solving it were proposed \cite{chertkov2018computational,greiter2018method,qi2019determininglocal,turkeshi2019entanglement,pakrouski2020automaticdesignof,inui2023inverse}.

In this work, we use the method proposed in Refs.\ \cite{chertkov2018computational,greiter2018method} to numerically find Hofstadter- and Kapit-Mueller-like parent Hamiltonians for the lattice Laughlin states on a finite-generation Sierpi\'{n}ski triangle and carpet with 27 and 64 sites, respectively. If a restriction to local hoppings is imposed, these Hamiltonians are approximate, but the ground states of some of them display significant overlap with a model Laughlin state. As the range of hoppings is increased, the Hamiltonians become almost exact. We also look for anyonic excitations, and in some cases, by lowering the number of particles and introducing pinning potentials, we obtain states with high overlaps with model states with quasiholes. Our analysis of finite-size effects on the model wavefunction shows that the considered systems are too small to completely separate the quasiholes during the braiding, and thus to demonstrate the fractional statistics. We also perform similar computations for the Hamiltonian models to investigate to what extent they reproduce the results obtained for the model wavefunction.

We start by recalling the expression for the Laughlin state on fractals in Sec.\ \ref{sec:state}. In Sec.\ \ref{sec:method} we explain the method of finding the parent Hamiltonians proposed in \cite{chertkov2018computational,greiter2018method} and the details of its implementation. Next, in Sec.\ \ref{sec:shortrange} we present the obtained local Hamiltonians and analyze the overlaps of their ground states with model states. Section \ref{sec:longrange} is devoted to the study of the nonlocal Hamiltonians, showing that their ground states can represent the lattice Laughlin states nearly perfectly. In Sec.\ \ref{sec:anyons}, we add the pinning potentials in order to trap anyons, and show that the resulting ground states have high overlaps with model wavefunctions describing anyons for some of the local Hamiltonians, and for nonlocal Hamiltonians the overlap becomes nearly perfect. In Sec.\ \ref{sec:braiding}, we analyze the effects of small lattice size on the process of braiding of anyons described by the model wavefunctions. Then, we check how well these results are reproduced in systems described by local and non-local Hamiltonians. Section \ref{sec:conclusions} concludes the article. The Supplementary Material \cite{supplementary2} contains the numerical values of parameters for Hamiltonians considered in this work (and a few more), as well as some other data files and more numerical results.

\section{The Laughlin state on fractals}\label{sec:state}
Lattice Laughlin states on arbitrary lattices embedded in two dimensions have been constructed \cite{nielsen2012laughlin,tu2014lattice} utilizing a connection to conformal field theory \cite{moore1991nonabelions}, and the models considered here on fractal lattices are particular instances of that construction. We consider $N$ sites, with positions $(x_j,y_j)$ denoted as the complex numbers $z_j=x_j+iy_j$. We fill the system with $M$ particles, and impose the hardcore condition, i.e. the occupation $n_j$ of the given site is 0 or 1.  We write the occupation number basis as $\ket{\mathbf{n}}$, where $\mathbf{n}=[n_1, n_2, \dots, n_N]$. The lattice analog of a Laughlin state with filling factor $1/q$ is given by
\begin{equation}
\ket{\Psi}=\frac{1}{C}\sum_{\mathbf{n}}\Psi_\mathbf{n}\ket{\mathbf{n}},
\label{eq:state}
\end{equation}
where $C$ is the normalization constant, and the coefficients $\Psi_\mathbf{n}$ are
\begin{multline}
\Psi_\mathbf{n}=\delta\left(q\sum_j n_j-N\eta \right)\prod_{j<k}(z_j-z_k)^{q n_j n_k} \\ \times
\prod_{j\neq k}(z_j-z_k)^{- n_j \eta},
\label{eq:laughlin} 
\end{multline}
where $\eta=qM/N$ is the flux per site and the Kronecker delta ensures the charge neutrality, i.e.\ the fact that we consider only the configurations $\mathbf{n}$ with the total number of particles $\sum_j n_j=M$.

One can also construct the wavefunctions for a lattice Laughlin state with anyons \cite{nielsen2015anyon}. In this work, we study localized quasiholes, each corresponding to a local particle density depletion of $1/q$ (i.e. introducing $q$ such quasiholes corresponds to removing one particle). We denote the position of the $l$th quasihole as $w_l\in \mathbb{C}$. This position can coincide with a given lattice site, but does not have to. Note that although we refer to $w_l$ as the ``anyon position'', the anyon itself is an extended object that lives on the sites of the fractal lattice as a local density depletion in the vicinity of $w_l$.

Let us start from a system with $M=M_0$ particles, described by \eqref{eq:laughlin}. After introducing $N_{\mathrm{qh}}$ quasiholes ($N_{\mathrm{qh}}/q\in\mathbb{N}$), the number of particles is $M=M_0-N_{\mathrm{qh}}/q$. The state $\ket{\tilde{\Psi}}$ with localized quasiholes is defined analogously to Eq. \eqref{eq:state}, with coefficients
\begin{multline}
\tilde{\Psi}_\mathbf{n}(\mathbf{w})=\delta\left(q\sum_j n_j+N_{\mathrm{qh}}-N\eta \right)
\\ \times
\prod_{j,l}(w_l-z_j)^{n_j}
\prod_{j<k}(z_j-z_k)^{q n_j n_k}
\\ \times
\prod_{j\neq k}(z_j-z_k)^{- n_j \eta}.
\label{eq:anyons} 
\end{multline}
The flux per site is still given by $\eta=qM_0/N=\frac{1}{N}\left(qM+N_{\mathrm{qh}}\right)$.

\begin{figure}
\includegraphics[width=0.45\textwidth]{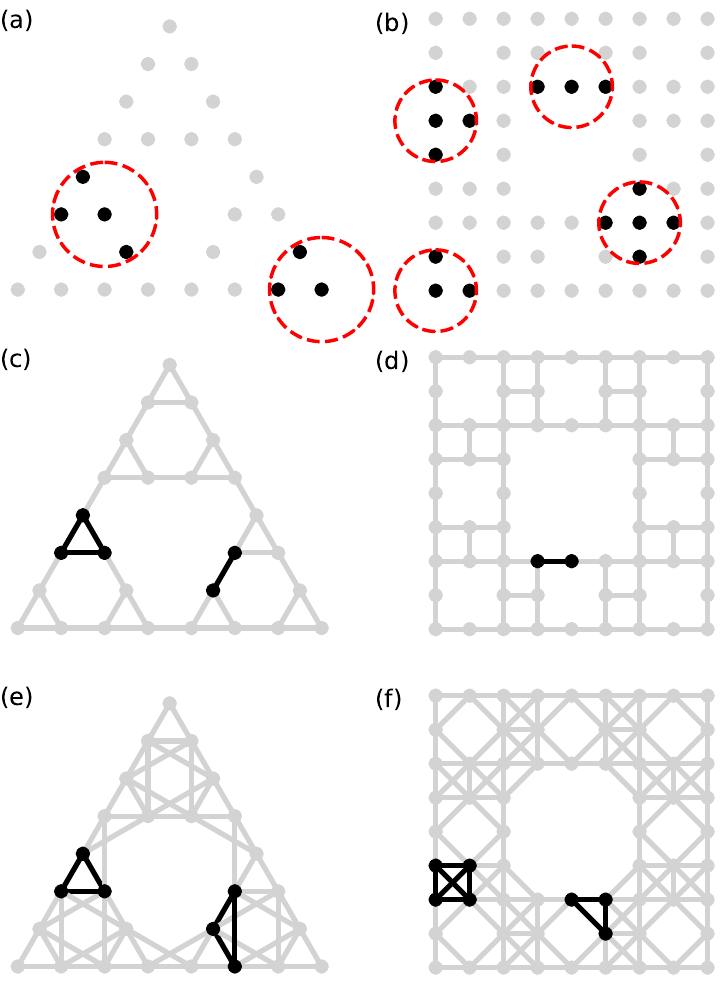}
\caption{The considered lattices: the 27-site Sierpi\'{n}ski triangle (left column) and the 64-site Sierpi\'{n}ski
carpet (right column). The rows show different choices for sets $S_\alpha$: the circles of radius $r=1.1$ around every site (top row), the NN-cliques (middle row) and the NNN-cliques (bottom row). The black dots denote the sites belonging to an example set, while the gray ones are all the other sites. In almost every variant, different sets of a given type can contain different numbers of sites -- thus several examples of sets are shown in these cases.
}
\label{fig:lattices}
\end{figure}

In the following, we limit ourselves to the case of $q=2$. We consider two examples of finite-generation fractal lattices: the Sierpi\'{n}ski triangle with $N=27$ sites and Sierpi\'{n}ski carpet with $N=64$ sites (see Fig. \ref{fig:lattices}). For brevity, in the following we will refer to these lattices as ``triangle'' and ``carpet'', respectively. Without loss of generality, we set the distance between nearest-neighboring sites to unity.

\section{The method for finding approximate parent Hamiltonians}\label{sec:method}

In this work, we look for hardcore boson tight-binding Hamiltonians whose ground state has a large overlap with \eqref{eq:laughlin}, that is, an approximate parent Hamiltonian of this state. We demand that the Hamiltonian has the following form
\begin{equation}
H=\sum_{j \neq k}t_{jk}e^{i\phi_{jk}}a^{\dagger}_j a_k+\sum_{j}\epsilon_j n_j,
\label{eq:hofstadter_general}
\end{equation}
where $a_j$ is an annihilation operator of a hardcore boson at site $j$, and $n_j=a_j^{\dagger}a_j$. The real parameters $t_{jk}$, $\phi_{jk}$ and $\epsilon_j$ are to be found. The hermiticity of the Hamiltonian requires that $t_{jk}=t_{kj}$ and $\phi_{jk}=-\phi_{kj}$. To enforce locality, one can consider $t_{jk}\neq 0$ only for e.g. nearest neighbours (NN), or nearest and next-to-nearest neighbours (NNN). To find $t_{jk}$, $\phi_{jk}$ and $\epsilon_j$, we use an ``inverse method'' \cite{greiter2018method,chertkov2018computational}. More specifically, we use one of the variants described in \cite{greiter2018method}. Below, we explain the method applied to our systems.

The idea is to find a set of (not necessarily Hermitian) approximate annihilation operators $A_{\alpha\beta}$ with $A_{\alpha\beta}\ket{\Psi}\approx 0$, where the target state $\ket{\Psi}$ in our case is the model wavefunction \eqref{eq:laughlin}. We construct the Hamiltonian as
\begin{equation}
H=\sum_{\alpha,\beta}A_{\alpha\beta}^{\dagger}A_{\alpha\beta}
\label{eq:h}
\end{equation}
In our case, a given operator $A_{\alpha\beta}$ will be constructed as linear combinations of the annihilation operators $a_j$ on a given cluster of sites located close to each other. We denote the set of site indices corresponding to operator $A_{\alpha\beta}$ as $S_\alpha=\{S_\alpha(1),S_\alpha(2),\dots S_\alpha(|S_\alpha|)\}$, where $|S_\alpha|$ is the size of the set $S_{\alpha}$, i.e. the number of included sites. A given site can belong to more than one cluster, (i.e.\ we can have e.g.\ $S_1=\{ 2,4,7\}, S_2=\{1,3,7\}$). For each set $S_{\alpha}$, we are going to construct several operators $A_{\alpha\beta}$, and the second index $\beta$ is introduced to differentiate between them. The choice of the site clusters $S_{\alpha}$ will determine the form of the resulting Hamiltonian. 

For more concreteness, let us look at the examples of clusters $S_{\alpha}$ which we will use in this work. One way to define the sets is to have one cluster $S_j$ assigned to each site $j$. The set contains the site $j$ and all the sites lying within the radius $r$ from it. Examples with $r=1.1$ are shown in Fig. \ref{fig:lattices} (a) and Fig. \ref{fig:lattices} (b) for the triangle and the carpet, respectively. In the former case, the sets contain three or four sites, in the latter -- three, four or five sites.

Another option is to represent the fractal lattice as a graph, where each pair of nearest-neighboring vertices/sites are connected by an edge. The clusters $S_\alpha$ can be defined as cliques on that graph, i.e. the sets that have the property that any pair of vertices is connected by an edge. More specifically, we choose the clusters $S_\alpha$ as maximal cliques, i.e. cliques that cannot be expanded by adding a further vertex. The example maximal cliques are shown in Fig.\ \ref{fig:lattices} (c), (d). In the case of the triangle (Fig.\ \ref{fig:lattices} (c)), they contain two or three sites, while for the carpet (Fig. \ref{fig:lattices} (d)) it is always two sites. We will call the cliques constructed in that way NN-cliques. This approach can be extended by considering maximal cliques on a graph where all nearest and next-nearest neighbors are connected by an edge. The example maximal cliques are shown in Fig.\ \ref{fig:lattices} (e), (f). We refer to this case as NNN-cliques.

As noted above, we look for operators $A_{\alpha\beta}$ being linear combinations of the annihilation operators $a_j$ on the sites belonging to the cluster $S_\alpha$, that is,
\begin{equation}
A_{\alpha\beta}=\sum_{\gamma=1}^{|S_\alpha|} c^{(\alpha\beta)}_{\gamma} a_{S_\alpha(\gamma)}, 
\label{eq:Acombination}
\end{equation}
where $c^{(\alpha\beta)}_{\gamma}$ are complex coefficients normalized to $\sum_{\gamma=1}^{|S_\alpha|} |c^{(\alpha\beta)}_{\gamma}|^2=1$. To obtain the coefficients $c^{(\alpha\beta)}_{\gamma}$, we apply the following procedure. We define the states $\ket{\phi^{(\alpha)}_\gamma}=a_{S_\alpha(\gamma)}\ket{\Psi}$ and the matrices $B^{(\alpha)}_{\gamma\delta}=\braket{\phi^{(\alpha)}_\gamma|\phi^{(\alpha)}_\delta}$. Then, we look for the eigenvectors of $\mathbf{B}^{(\alpha)}$,
\begin{equation}
\sum_\delta B^{(\alpha)}_{\gamma\delta} c^{(\alpha\beta)}_{\delta}=\lambda^{(\alpha\beta)} c^{(\alpha\beta)}_{\gamma},
\label{eq:Beigenproblem}
\end{equation}
where $\lambda^{(\alpha\beta)}$ is the $\beta$th eigenvalue of $\mathbf{B}^{(\alpha)}$, and $c^{(\alpha\beta)}_{\delta}$ are the complex coefficients of the $\beta$th eigenvector. To construct the annihilation operators, we choose the eigenvectors corresponding to nearly-zero eigenvalues $\lambda^{(\alpha\beta)}$. We use the coefficients $c^{(\alpha\beta)}_{\delta}$ of these selected eigenvector as the expansion coefficients in \eqref{eq:Acombination}.

How do we quantify the closeness of $\lambda^{(\alpha\beta)}$ to zero? That is, how do we determine how many operators $A_{\alpha\beta}$ to construct and which eigenvectors of $\mathbf{B}^{(\alpha)}$ to use as expansion coefficients? One option, which we use in most of the cases, is to set a threshold $d$, and use only the eigenvectors corresponding to $\lambda^{(\alpha\beta)}<d$. The other option is to set a number $m$, and use $m$ eigenvectors corresponding to the lowest $\lambda^{(\alpha\beta)}$ for each cluster $S_\alpha$. In this work, we only use $m=1$.

The Hamiltonian \eqref{eq:h} has the form \eqref{eq:hofstadter_general} by construction. The hardcore constraint is imposed implicitly by using the subspace of the Hilbert space which fulfills the constraint. The range of the hoppings in \eqref{eq:h} is determined by the chosen sets $S_\alpha$: the result of the multiplication of $A^{\dagger}_{\alpha\beta}$ and $A_{\alpha\beta}$ are hopping terms connecting pairs of sites $(j,k)$ with $j,k\in S_\alpha$, and the onsite potentials on sites $j \in S_\alpha$. That is, the $r=1.1$, NN-clique and NNN-clique cases presented in Fig.\ \ref{fig:lattices}, lead to up to third-, first- and second-neighbor hoppings, respectively. We note that in our calculations, the conversion from $A_{\alpha\beta}$ to $t_{jk}$, $\phi_{jk}$ and $\epsilon_{j}$ is approximate, i.e. contributions smaller than a certain threshold are discarded.

In addition to the approach outlined above, there are several alternative variants of the method described in \cite{greiter2018method,chertkov2018computational}. For example, instead of the terms of the annihilation operator, one can look for the terms of the Hamiltonian itself. We have not used this method here, however, since in our case it leads to non-Hermitian Hamiltonians because the Laughlin state is complex. Constructing the Hamiltonian using annihilation operators ensures that it is both Hermitian and positive semi-definite.

Another option mentioned in \cite{greiter2018method} is to refine the result by optimizing e.g.\ energy variance or overlap using numerical methods such as the Newton scheme, although the authors of \cite{greiter2018method} find that this is not necessary in most cases, and it is more demanding numerically. We therefore do not do this here. We note that there are also other methods, such as the entanglement-guided approach \cite{turkeshi2019entanglement} or direct optimization of a cost function composed of several observables \cite{pakrouski2020automaticdesignof}, which may be useful in the further study of topological orders on fractals.

\begin{figure}
\includegraphics[width=0.45\textwidth]{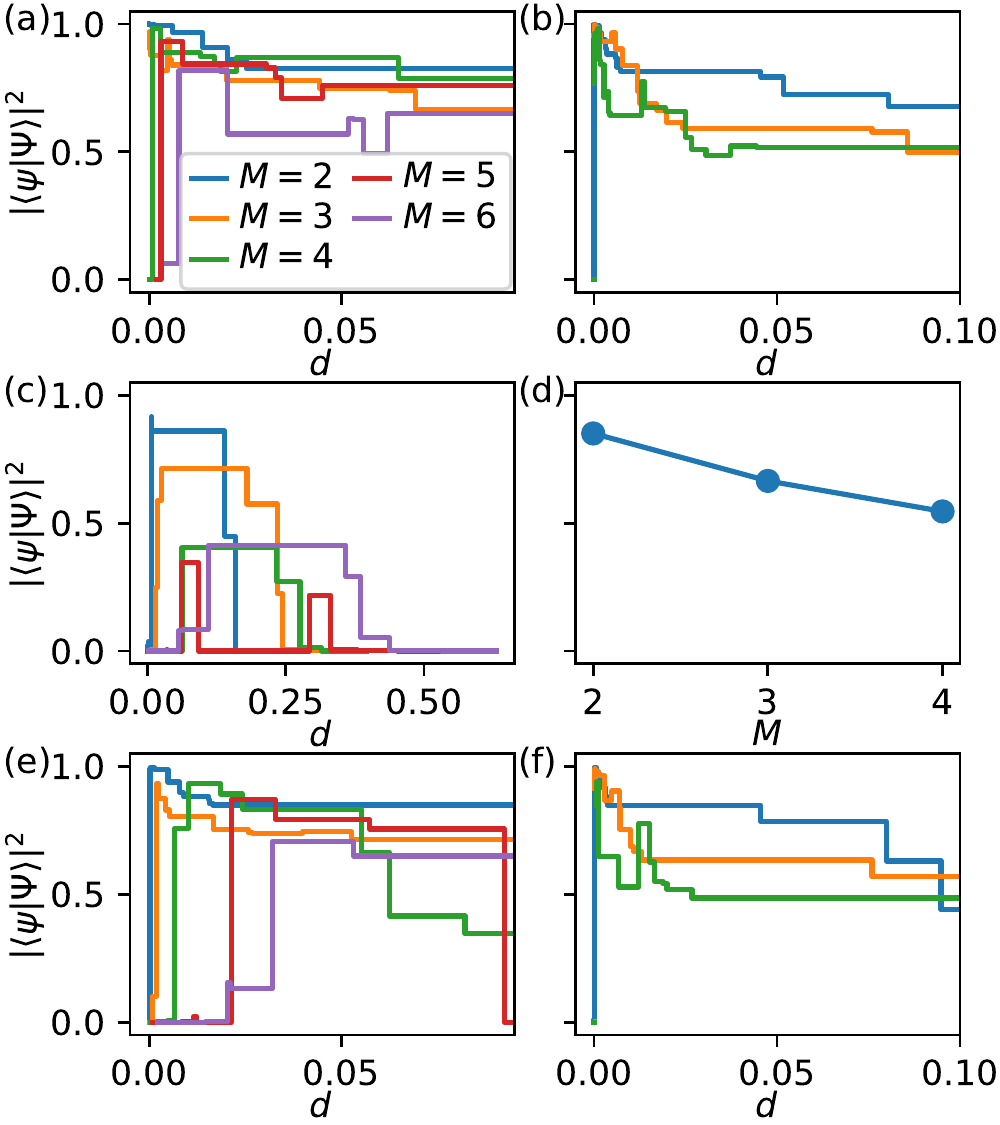}
\caption{The squared overlaps between the Laughlin state \eqref{eq:laughlin} and the ground state of the considered Hamiltonians for the fractal lattices with up to third neighbor hoppings. All the plots show the overlap vs.\ threshold $d$, except from (d) which displays overlap vs.\ system size. The subplots correspond to the choices of $S_\alpha$ shown in Fig. \ref{fig:lattices}. That is, the left (right) column corresponds to the triangle (carpet), and the top, middle, and bottom row corresponds to the choices: $r=1.1$, NN-cliques, NNN-cliques, respectively. In all subfigures but (d) the colors correspond to different particle numbers $M$.
}
\label{fig:OverlapsVsThresholds}
\end{figure}

\begin{figure}
\includegraphics[width=0.45\textwidth]{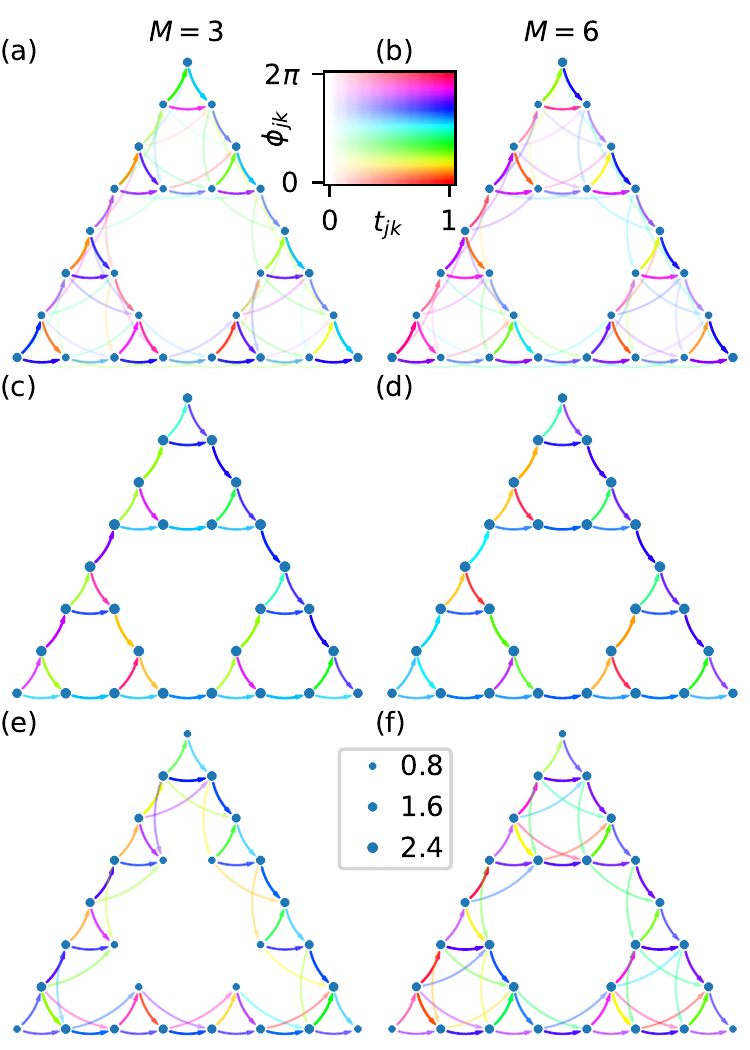}
\caption{Examples of the Hamiltonians for the triangle. Hoppings are denoted by the arrows, with brightness representing the modulus and color representing the phase (see the inset). The dots represent the onsite potentials, with size representing the strength (see the legend). The terms are renormalized so that for each Hamiltonian the strength of the strongest hopping is unity. The columns correspond to the particle number: (a), (c), (e) $M=3$, (b), (d), (f) $M=6$. The rows correspond to types of site clusters: (a), (b)  $r=1.1$, (c), (d) NN-cliques, (e),(f) NNN-cliques.
}
\label{fig:HamiltoniansTriangle}
\end{figure}

\begin{figure}
\includegraphics[width=0.45\textwidth]{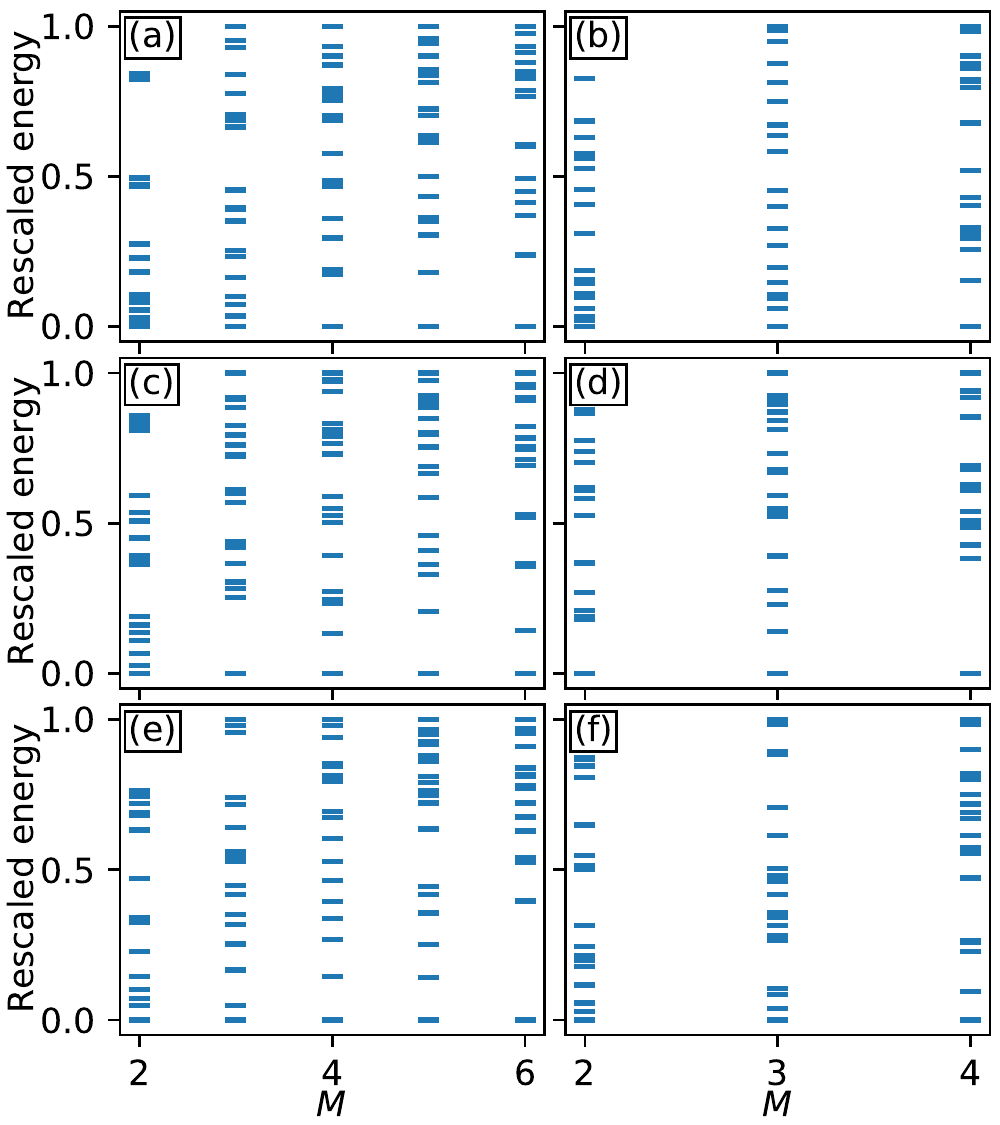}
\caption{The low-energy spectra of the Hamiltonians. The subplots are organized in the same way as in Fig.\ \ref{fig:lattices}. That is, the left (right) column corresponds to the triangle (carpet), and the top, middle, and bottom row correspond to the choices: $r=1.1$, NN-cliques, NNN-cliques, respectively. The energy is rescaled so that the 1st eigenvalue has energy 0 and the 20th eigenvalue has energy 1.
}
\label{fig:spectra}
\end{figure}

\section{Results: local Hamiltonians}\label{sec:shortrange}

In this section, we use the method from Sec.\ \ref{sec:method} to obtain Hamiltonians with up to third-neighbor hopping, with $S_\alpha$ chosen either as $r=1.1$, NN-cliques or NNN-cliques. We consider the triangle with the number of particles $M$ from 2 to 6, and the carpet for $M=2,3,4$. After obtaining the Hamiltonian, its ground state is compared to the target state, i.e.\ the model state \eqref{eq:laughlin} with $M$ particles and flux per site $\eta=qM/N$. Throughout this work, all the ground states and low-energy spectra are found by employing the exact diagonalization method for sparse matrices, implemented using the ARPACK library (except from single-particle spectra in Sec.\ \ref{sec:longrange}, which are obtained by diagonalizing dense matrices using the LAPACK library).

In Figure \ref{fig:OverlapsVsThresholds}, we show the squared overlap $|\braket{\psi|\Psi}|^2$ between the ground state $\ket{\psi}$ of the Hamiltonian and the model state $\ket{\Psi}$ defined by Eq.\ \eqref{eq:laughlin} for different systems. In most of the subfigures, squared overlap is plotted as a function of the threshold $d$ for different systems. The only exception is Fig.~\ref{fig:OverlapsVsThresholds}~(d), which corresponds to the NN-clique case on the carpet, where the $\mathbf{B}^{(\alpha)}$ matrix can only have size $2\times 2$, so instead of applying a threshold $d$ we only take the lowest eigenvalue. Thus, instead of plotting the squared overlap vs.\ $d$, we plot the squared overlap vs.\ $M$.

The plots of $|\braket{\psi|\Psi}|^2$ vs.\ $d$ in Fig.~\ref{fig:OverlapsVsThresholds} exhibit a number of plateaus. This is because, although $d$ is a continuous parameter, the eigenvalues of the density matrices are discrete, thus any $d$ between two nearest eigenvalues will yield the same Hamiltonian.

The right choice of $d$ is important to get a good parent Hamiltonian. Too small $d$ would mean that in some clusters, no eigenvalue $\lambda^{(\alpha\beta)}$ would fulfill $\lambda^{(\alpha\beta)}<d$. As a consequence, some hoppings will be absent, which will lead to small squared overlaps $|\braket{\psi|\Psi}|^2$. In contrast, too large $d$ means that the included eigenvalue will be far from zero, and $A_{\alpha\beta}$ would no longer behave as an approximate annihilation operator, which also leads to small overlaps. Indeed, in the plots in Fig.~\ref{fig:OverlapsVsThresholds}, the maximum $|\braket{\psi|\Psi}|^2$ is seen at small, but not too small, $d$ (the overlap eventually falls to almost zero for high enough $d$, which is not seen in most of the plots, which show only low $d$). In the following, when mentioning the Hamiltonian for a given system, we mean the Hamiltonian for a $d$ yielding maximum overlap (except the case of NN-cliques on the carpet, where this would mean the Hamiltonian generated from the lowest eigenvalue $\lambda^{(\alpha\beta)}$ of each cluster).

From Fig.~\ref{fig:OverlapsVsThresholds} (a),(b) one can see that the $r=1.1$ case, corresponding to the third-neighbor hoppings, yields squared overlaps above 0.81 for all particle numbers and both lattices, and above 0.96 for both lattices with $M\leq 4$. In the case of the NNN-cliques (i.e. second-neighbor hoppings), seen in Fig.~\ref{fig:OverlapsVsThresholds} (e),(f), the squared overlaps are smaller, but still above 0.87 for both lattices and all the considered $M$ values, except from the triangle with $M=6$, for which $|\braket{\psi|\Psi}|^2\approx 0.71$. For the NN-cliques (Fig.~\ref{fig:OverlapsVsThresholds} (c),(d)), the overlaps are much worse, exceeding 0.8 only for $M=2$ (both lattices), and reaching as low as $|\braket{\psi|\Psi}|^2\approx 0.35$ for $M=5$ on the triangle.

Examples of the resulting Hamiltonians for the triangle are shown in Fig.\ \ref{fig:HamiltoniansTriangle}. Here, the arrows denote the hoppings, with brightness and color representing the strength and phase, respectively, for the hopping in the arrow direction (in the opposite direction the coefficient is the complex conjugate). To obtain clear plots with a common color scale, the plotted terms of the Hamiltonians are normalized, so that the absolute value of the strongest hoppping in each of them is 1. Other results for both the triangle and the carpet, as well as the numerical values of Hamiltonian parameters, can be found in the Supplementary Material \cite{supplementary2}.

In the particular case of NNN-cliques with $M=3$ on the triangle, some nearest-neighbor hoppings are absent, which makes the system equivalent to a fourth-neighbor one-dimensional model. The resulting Hamiltonian is plotted in Fig \ref{fig:HamiltoniansTriangle} (e). In all the other cases, these hoppings are present.

The spectra of the Hamiltonians are shown in Fig.\ \ref{fig:spectra}. The plot contains the 20 lowest energy eigenvalues, with energy rescaled so that the lowest and highest one have energies 0 and 1, respectively. In some cases, but not all, we observe an energy gap above the ground state, as seen for lattice quantum Hall systems with hard-wall boundary conditions \cite{glasser2015exact}. By energy gap above the ground state, we here mean that the energy difference between the ground state and the first excited state is significantly larger than the energy differences among the lowest excited states.

In summary, we have found local Hamiltonians (i.e. ones with up to third-neighbor hoppings) whose ground states have a reasonable overlap with model Laughlin states on the triangle and the carpet. That is, for each studied value of $M$ on each lattice, we found at least one Hamiltonian with ground state fulfilling $|\braket{\psi|\Psi}|^2>0.81$.

\begin{figure}
\includegraphics[width=0.5\textwidth]{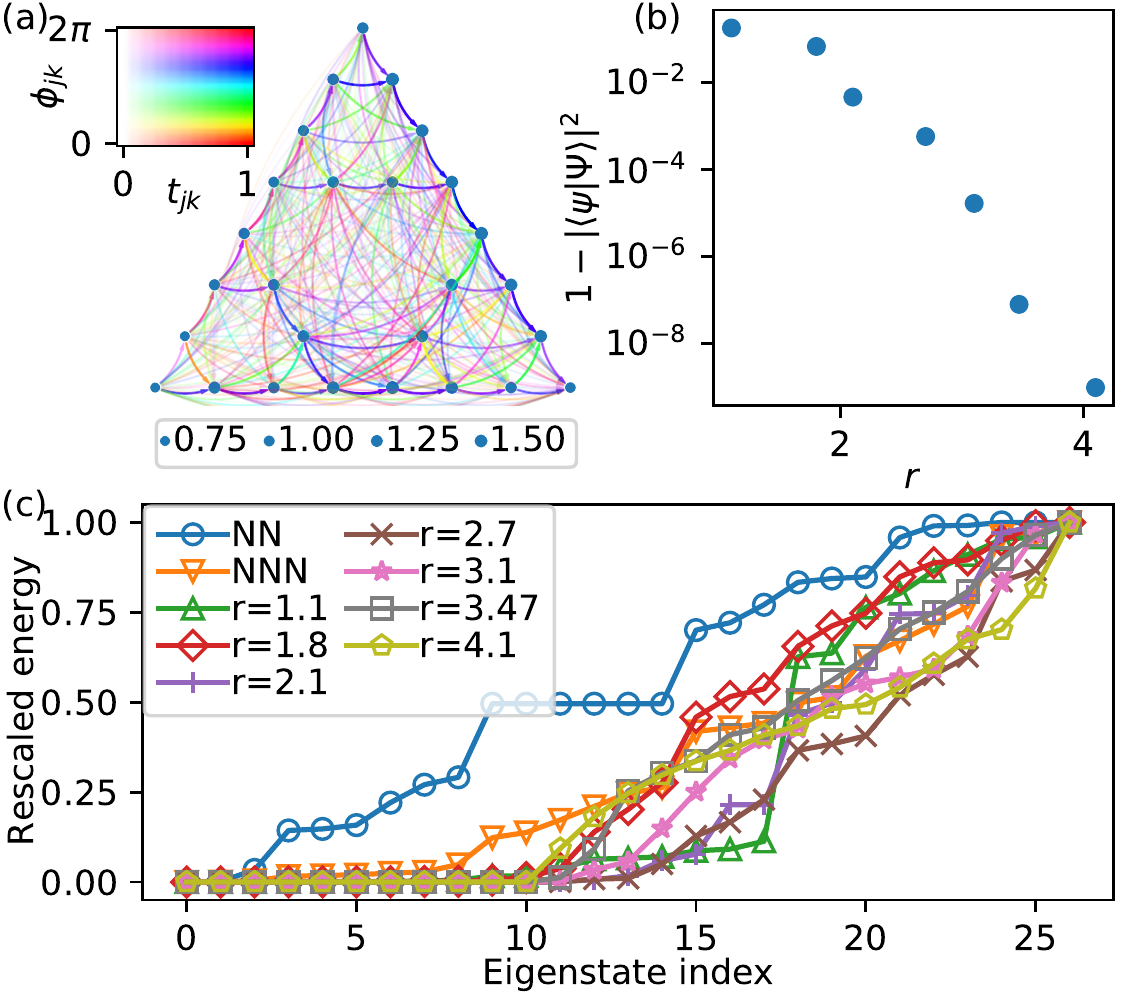}
\caption{The properties of the nonlocal Hamiltonians constructed for an $M=6$ target state. (a) The hoppings and onsite potentials of the $r=4.1$ Hamiltonian. (b) The error of the approximation of the model state by the ground state of the Hamiltonians, as a function of radius $r$. The error is defined as one minus the squared overlap. (c) The single-particle energy spectra of various Hamiltonians, showing the emergence of a flat ``band'' for high $r$.}
\label{fig:longrange}
\end{figure}

\section{Results: nonlocal Hamiltonians}\label{sec:longrange}

The range of the hoppings can be easily increased by increasing the radius $r$. Figure \ref{fig:longrange} (a) displays an $r=4.1$ Hamiltonian for the triangle with $M=6$ (see also \cite{supplementary2} for the numerical values of the parameters of this and other Hamiltonians considered in this section). In such a case the hoppings span almost across the whole lattice. The overlap of the $M=6$ ground state with the model state \eqref{eq:laughlin} is almost perfect, with $1-|\braket{\psi|\Psi}|^2<10^{-8}$. 

In Fig.\ \ref{fig:longrange} (b), we plot the ``error'' $1-|\braket{\psi|\Psi}|^2$ for the $M=6$ case on a triangle, with $r$ increasing from $r=1.1$ to $r=4.1$. It can be seen that the overlap gradually approaches 1 when $r$ grows.

It can be instructive to look at single-particle spectra of the studied Hamiltonians. That is, we construct the Hamiltonians with the model Laughlin wavefunction at given $M$ as a target state, and then, without changing their parameters, we diagonalize them at $M=1$. The results for Hamiltonians created for an $M=6$ triangle are shown in Fig.\ \ref{fig:longrange} (c). For the sake of comparison between different systems, the energy is rescaled so that the first eigenvalue is zero and the last eigenvalue is unity. It can be seen that as the maximum hopping distance increases, a nearly-flat ``band'' of lowest-energy states forms, reminiscent of a Landau level in continuum two-dimensional systems. For $r=4.1$, this ``band'' contains 11 states. This is similar to the case of a Landau level on a disk or a cylinder. In such systems, a Laughlin state of $M$ particles without anyons is composed out of $q(M-1)+1$ single-particle orbitals (see e.g.\ \cite{mitra1993angular}). If we take $M=6$, $q=2$, then $q(M-1)+1=11$, which suggests that we can treat the 11 lowest-energy single-particle states of the nonlocal Hamiltonians as analogs of the Landau level orbitals.

Thus, the nonlocal Hamiltonian is similar to the Kapit-Mueller model, a two-dimensional lattice model which also has complex hoppings with arbitrary hopping distance. The Kapit-Mueller model has an exactly flat band, spanned by lattice analogs of lowest Landau level wavefunctions, allowing to exactly realize discretized Laughlin wavefunctions \cite{kapit2010exact}.

\begin{figure}
\includegraphics[width=0.45\textwidth]{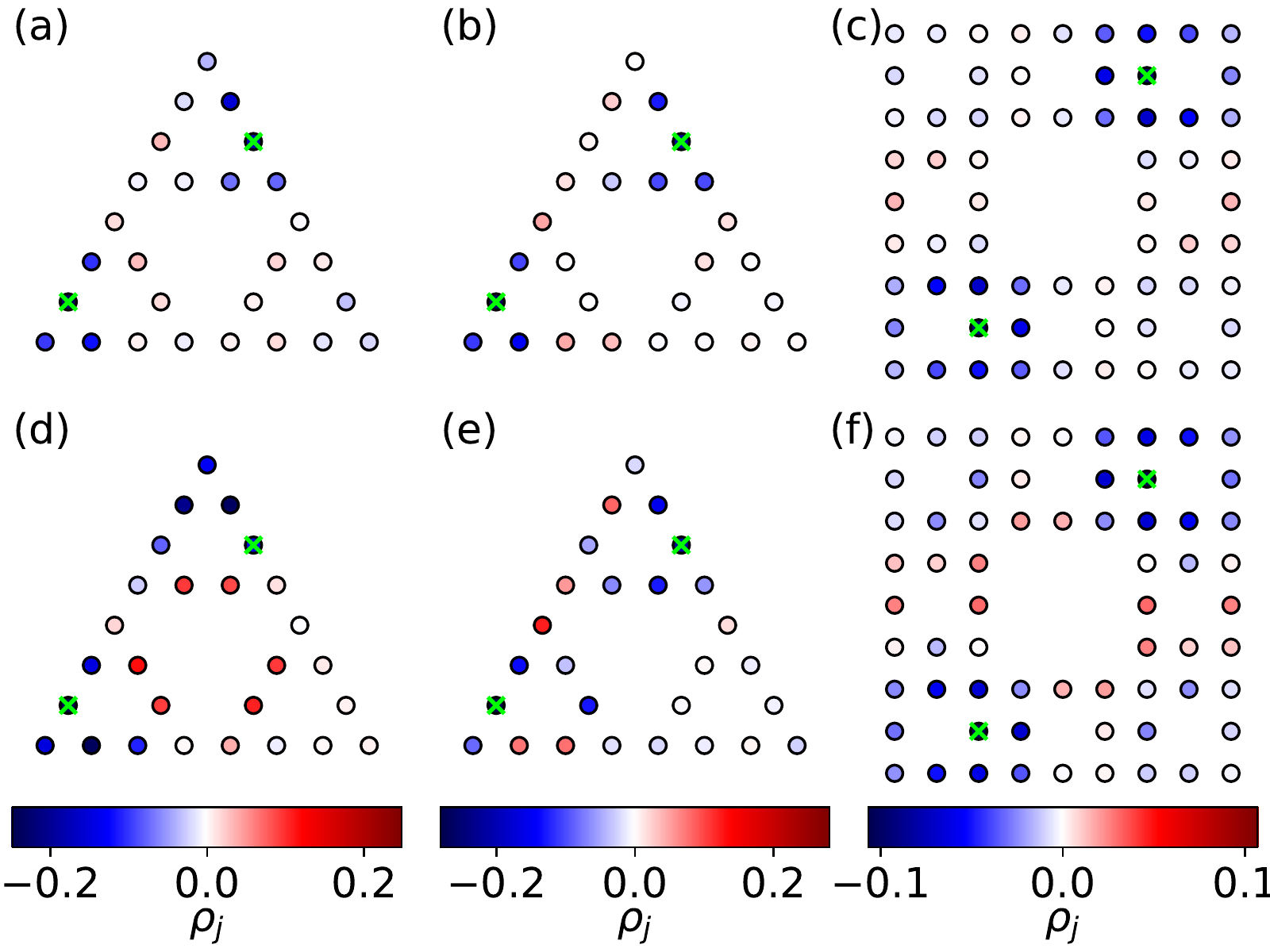}
\caption{The plots of excess particle density \eqref{eq:excesscharge} for six example Hamiltonians. The positions of the potentials are denoted by bright green crosses. The results are shown for the following cases: 
(a) triangle $M_0=5$, $r=1.1$,  (b) triangle, $M_0=6$, $r=4.1$, (c) carpet, $M_0=4$, $r=1.1$, 
(d) triangle, $M_0=5$, NN-cliques, (e) triangle, $M_0=6$, $r=1.1$, (f) carpet, $M_0=4$, NN-cliques. Each colorbar in the bottom corresponds to both plots in a given column.}
\label{fig:DensityDifference}
\end{figure}

\begin{figure}
\includegraphics[width=0.5\textwidth]{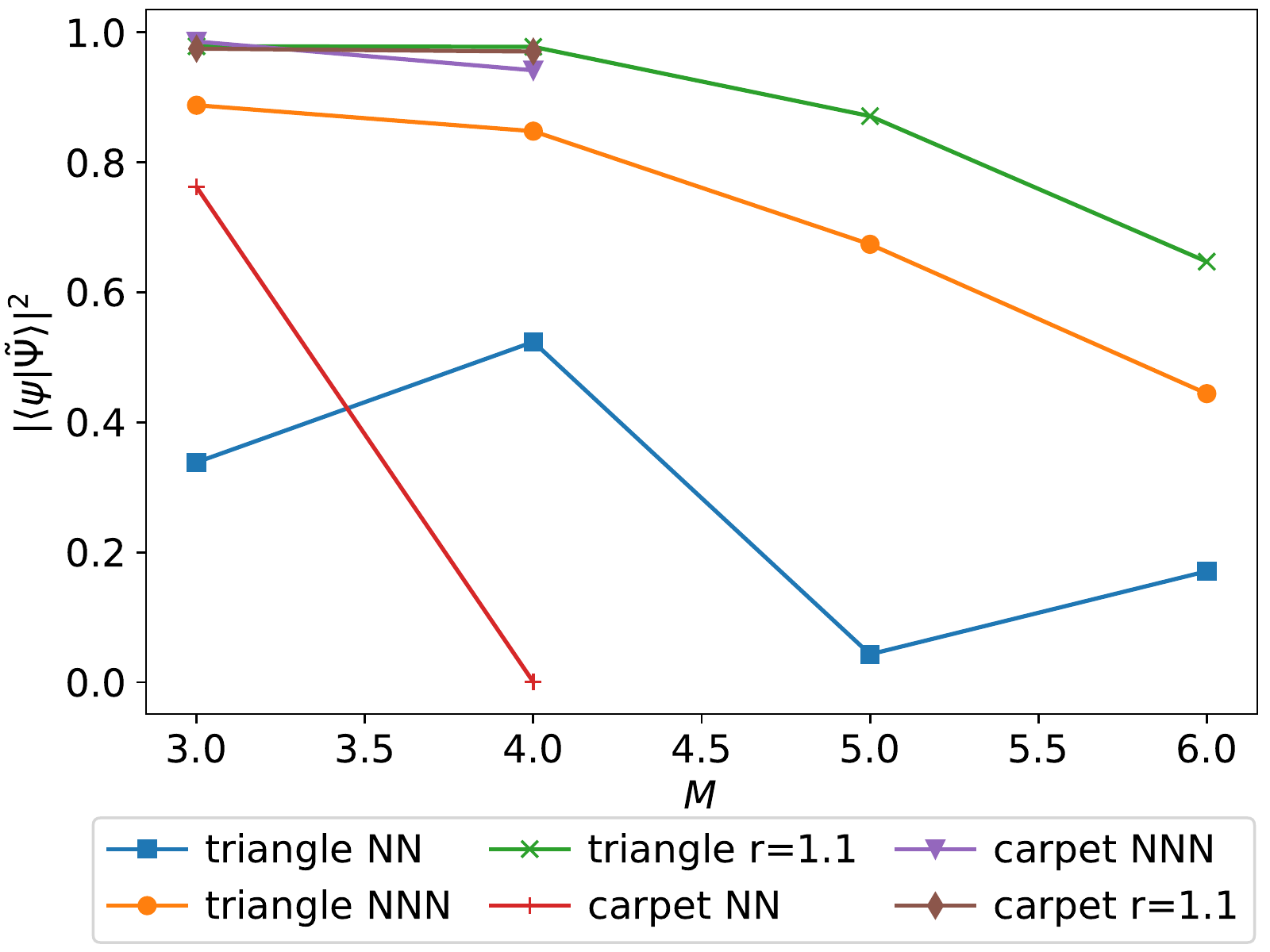}
\caption{The squared overlaps of the ground states of our models at $M=M_0-1$ with model quasihole states \eqref{eq:anyons}. In the exact diagonalization, we use two potentials located as in Fig.\ \ref{fig:DensityDifference} ($V=1000$). The anyon positions in the model wavefunction are the same as the positions of the potentials in the Hamiltonian models.}
\label{fig:OverlapQuasiholes}
\end{figure}

\section{Results: anyons}\label{sec:anyons}

One of the crucial characteristics of topological orders is the presence of anyonic excitations, which, for the quantum Hall systems, have the form of quasielectrons and quasiholes. Here, we attempt to create localized quasiholes in the Hamiltonian models constructed in Secs.\ \ref{sec:shortrange} and \ref{sec:longrange}.

To avoid confusion, we denote the particle number $M$ of the target model state \eqref{eq:laughlin}, for which the Hamiltonian was generated, as $M_0$, while the symbol $M$ throughout this section and Sec.\ \ref{sec:braiding} will denote the particle number with which we work at the moment. In general, we can have $M_0\neq M$. The Hamiltonian parameters are set by $M_0$ and do not depend on $M$. 

In an attempt to create two quasiholes in the system described by a Hamiltonian $H$, constructed for a target model state with $M_0$ particles, we remove one particle from the system, i.e.\ we set $M=M_0-1$. Then, we add two onsite potentials, which are supposed to trap them (i.e.\ the total Hamiltonian is $H'=H+Vn_a+Vn_b$, where $a$ and $b$ are the two chosen sites, and we choose $V=1000$). If the system hosts anyons, then the excess particle density of $-1/2$ should be located in the vicinity of each of the trapping potentials -- although we note that the anyons have a finite extent, and the system may be too small for them to be well separated.

Fig.\ \ref{fig:DensityDifference} shows the plots of excess particle density distribution for several example cases. The excess particle density is defined as 
\begin{equation}
\rho_j=\braket{n_j}_{M_0-1}-\braket{n_j}_{M_0},
\label{eq:excesscharge}
\end{equation}
where the $\braket{n_j}_{M_0}$ ($\braket{n_j}_{M_0-1}$) is the expectation value of particle density at site $j$ in the ground state with $M=M_0$ without additional potentials ($M=M_0-1$ with additional potentials). In some of the studied cases, for example the ones shown in Fig. \ref{fig:DensityDifference} (a), (b), (c), one can see that the strongest density depletion (dark blue) is located in the vicinity of the pinning potentials, resembling particles with well-defined position. However, there are some smaller excess density variations even far away from the pinning potentials, suggesting that the anyons (if they are indeed anyons) are too big to be completely separated within the structure. Also, we note that in some of the studied cases, such as in Fig. \ref{fig:DensityDifference} (d), (e), (f), these variations are stronger, and the plots are less reminiscent of pinned quasiparticles.

In general, the cases with only nearest-neighbor hopping fail to produce well-localized density depletions (see Fig. \ref{fig:DensityDifference} (d) and (f)). The localization improves when the range of the hopping is increased (compare e.g. Fig. \ref{fig:DensityDifference} (a) and (d)). Also, as $r$ increases, the excess particle density distribution approaches the distribution obtained for the model wavefunctions \eqref{eq:laughlin} and \eqref{eq:anyons}. At sufficiently high $r$ (e.g. Fig. \ref{fig:DensityDifference} (b)), these two match almost perfectly.

The results depend also on the particle number, although the dependence is not straightforward: for the $r=1.1$ Hamiltonians on the triangle, particularly good results are achieved for the $M_0=5$, $r=1.1$ case, which is neither the highest nor the lowest considered $M_0$ (Fig. \ref{fig:DensityDifference} (a)). We speculate that the influence of particle number can be twofold. First, based on Fig.~\ref{fig:OverlapsVsThresholds}, we expect that higher $M_0$ means that the system represents the Laughlin physics less accurately. But secondly, the size of the anyons described by \eqref{eq:anyons} decreases when we increase $\eta$ (at least at sufficiently small $\eta$), which happens when we increase $M_0$ and keep $N$ constant. Obviously, for sufficiently nonlocal Hamiltonians whose ground states faithfully represent the excess particle density distribution of model states, $\eta$ becomes the only factor, and thus we expect that the anyon size will decrease with $M_0$.

For quantitative assessment of how well our models are suited to host anyons, we compute the overlaps  between the ground states $\ket{\psi}$ with potentials and $M=M_0-1$ particles, and the model wavefunctions with two anyons \eqref{eq:anyons}. This method is not perfect, as excitations can have anyonic statistics even when they are not described by \eqref{eq:anyons}. Nevertheless, because our models are designed to generate a ground state approximating \eqref{eq:laughlin} at $M_0$ particles, \eqref{eq:anyons} is a reasonable guess for $M_0-1$ particles. The resulting squared overlaps $|\braket{\psi|\tilde{\Psi}}|^2$ for local models are shown in Fig.\ \ref{fig:OverlapQuasiholes}. It can be seen that the overlaps seem to depend on two factors: the number of particles and the range of the hoppings. Both are understandable. With a given type of site cluster (e.g. $r=1.1$) and lattice, the overlaps with no anyons (Fig.\ \ref{fig:OverlapsVsThresholds}) are highest for small number of particles, so it is not surprising that these cases also yield best overlaps for states with quasiholes (also, such systems have the smallest Hilbert spaces). The overlaps also grow with increasing hopping range, as in such cases the ground state at $M=M_0$ particles represents \eqref{eq:laughlin} better.  In general, for $M_0=3,4$ and $M=M_0-1$ the $r=1.1$ models yield squared overlaps over $|\braket{\psi|\tilde{\Psi}}|^2>0.95$ on both lattices.

Even better overlaps can be achieved by increasing the range $r$. For large enough $r$, the overlap can be made almost equal to 1. For example $1-|\braket{\psi|\tilde{\Psi}}|^2<10^{-8}$ for the $r=4.1$ model on the triangle with $M_0=6$. We note that the exact Hamiltonians for systems with anyons known from conformal field theory are constructed from the no-anyon Hamiltonians by modifying the coefficients of terms involving all the sites \cite{nielsen2015anyon}, while in our case we obtain an almost-exact parent Hamiltonian by modifying only two onsite potentials.

\begin{figure}[h!]
\includegraphics[width=0.45\textwidth]{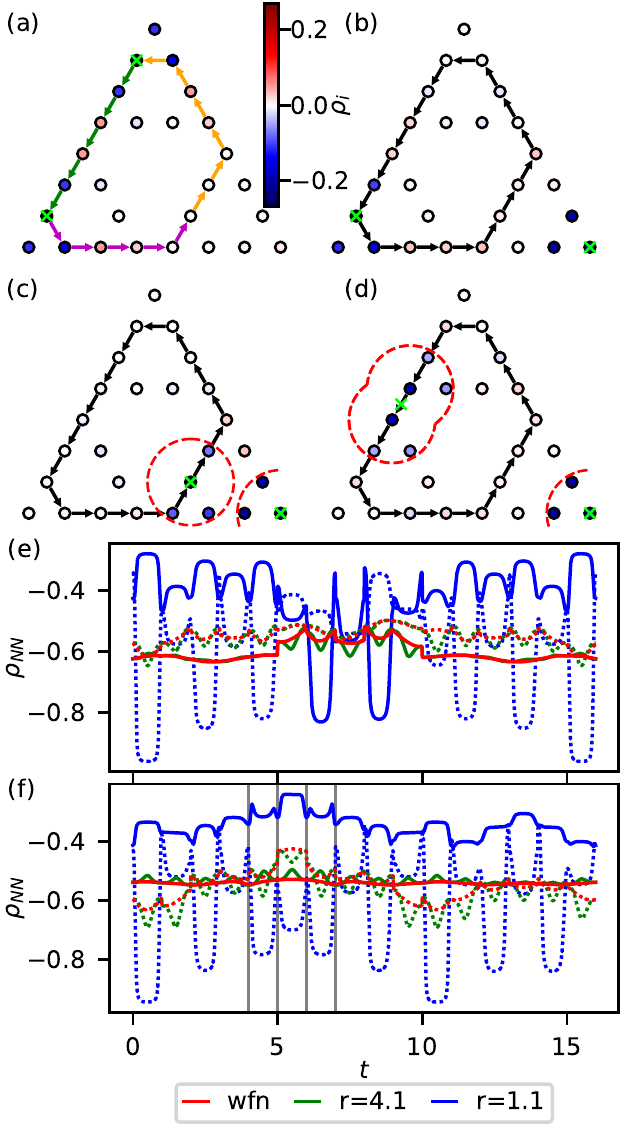}
\caption{Braiding in a triangle with $M_0=6$ and $M=5$. (a)-(d) show the excess particle densities \eqref{eq:excesscharge} of the model wavefunction: (a) at the beginning of the exchange path, (b) at the beginning of the AB path, (c) at a point of the AB path when the two anyons seem to blend with each other, (d) at some point of the AB path where the anyon is located halfway between two sites. The bright green crosses in (a)-(d) denote the positions $w_k$ of the anyons. The red dashed shapes enclose sites at a distance at most 1 from $w_k$ (if $w_k$ coincides with a site) or from either of the two sites between which it is interpolated (if it does not). The arrows denote the paths of anyon motion. (e), (f) The local excess particle density throughout the exchange and AB paths, respectively. The dashed and solid lines correspond to $\rho_{\mathrm{NN}1}(t)$ and $\rho_{\mathrm{NN}2}(t)$, respectively, while different colors denote different cases (the model wavefunction, the $r=4.1$ Hamiltonian and the $r=1.1$ Hamiltonian).  See the main text for more details.}
\label{fig:braiding_path1}
\end{figure}

\section{Results: braiding}\label{sec:braiding}

In sufficiently big fractal systems, the quasiholes described by the model wavefunction \eqref{eq:anyons} obey the same statistics as the quasiholes of the Laughlin wavefunctions \cite{manna2020anyon}. However, to clearly observe the statistical phase, the anyons need to be sufficiently separated, which may not be possible on small lattices. In the following, we will study how the small system size affects the braiding process for the model wavefunction \eqref{eq:anyons}. Then, we will compare these results to the results for systems described by Hamiltonians. 

We choose to focus on the case of $M_0=6$ on a triangle, where, because of relatively high $\eta$, the anyons seem to be quite small. The anyon positions $w_k$ are external parameters of the wavefunction, which can be varied continuously. Let us consider a closed path, parametrized as $\mathbf{w}(t)$, where $t\in[0,t_{\mathrm{max}}]$. We denote the wavefunction \eqref{eq:anyons} on the path as $\ket{\tilde{\Psi} (\mathbf{w}(t))}$. In numerical calculations, we consider a discretized version of the path at points $t_1, t_2,\dots, t_{N_{\mathrm{path}}}$. Then, an approximation to the Berry phase is given by
\begin{multline}
\gamma=-\mathrm{Im} \log \Biggl[ \left(\prod_{j=1}^{N_{\mathrm{path}}-1} \braket{\tilde{\Psi} (\mathbf{w}(t_j))|\tilde{\Psi} (\mathbf{w}(t_{j+1}))} \right) \times \\ \times
\braket{\tilde{\Psi} (\mathbf{w}(t_{N_{\mathrm{path}}}))|\tilde{\Psi} (\mathbf{w}(t_{1}))} \Biggr]
\end{multline}
\cite{resta2000manifestations}. The quality of this approximation increases with increasing number of discretized points.
Figure \ref{fig:braiding_path1} (a) shows the exchange path considered by us. We start by placing the anyons on two sites, and then move one anyon at a time between two nearest-neighboring sites. The anyon positions between the sites are interpolated linearly with $n_{\mathrm{interp}}$ steps. The anyons are moved in the following way: first, one anyon moves along the purple arrows, then the second anyon moves along the green arrows, and finally the first anyon moves along the orange arrows. In this way, $w_1$ gets transformed into $w_2$ and vice versa, i.e. the positions of the anyons are exchanged. 

More formally, we consider $t=T+s$, where $T\in \mathbb{N}^0$, and $s\in [0,1)$ (i.e. $T=\lfloor t \rfloor $, and $s=(t\mod 1)$). To define the path, we introduce two sequences of site indices: $k(T)$ and $l(T)$. At integer $t$ (i.e.\ $t=T$), the anyons are located at sites: $w_1(t)=z_{k(T)}$ and $w_2(t)=z_{l(T)}$. At noninteger $t$, we interpolate between the sites: $w_1(t)=(1-s)z_{k(T)}+sz_{k(T+1)}$, and analogously for $w_2$. Because we only move one anyon at a time, we define the sequences $k(T)$ and $l(T)$ in such a way that at each $T$, we either have $k(T)=k(T+1)$ or $l(T)=l(T+1)$. When we discretize the path for numerical calculation, we divide the interpolation into $n_{\mathrm{interp}}$ equal steps, i.e. $t_j=T_j+s_j$, with $s_j=m_j/n_{\mathrm{interp}}$, and $m_j\in {0,1, \dots, n_{\mathrm{interp}}-1}$.

The phase $\gamma_{\mathrm{exc}}$ on the path shown in Fig. \ref{fig:braiding_path1} (a) contains both the statistical phase and the Aharonov-Bohm (AB) phase. To determine the AB contribution, we consider a situation shown in Fig.\ \ref{fig:braiding_path1} (b): one anyon goes around the path denoted by black arrows, and the other one is located at a constant position outside of it (i.e. $l(T)=\mathrm{const}$), in the corner of the triangle. The resulting phase $\gamma_{\mathrm{AB}}$ is then subtracted from $\gamma_{\mathrm{exc}}$ to obtain the braiding phase $\gamma_{\mathrm{br}}=\gamma_{\mathrm{exc}}-\gamma_{\mathrm{AB}}$. 

During the Aharonov-Bohm phase calculation, the anyons get particularly close to each other. In Fig.\ \ref{fig:braiding_path1} (c), one can see an excess density distribution for a particular part in the path, where the anyons seem to merge with each other.

We note that when anyons are far away from each other and each one is located at a given site, e.g.\ like in Fig.\ \ref{fig:braiding_path1} (b), the majority of the density depletion is located at that site and its nearest neighbors. If we approximate the anyons as objects of radius 1, they can be regarded as separated even in the situation from Fig.\ \ref{fig:braiding_path1} (c) -- see the dashed red circles. In the case of anyon located in between two sites (like the case presented in Fig.\ \ref{fig:braiding_path1} (d), where $s=0.5$) we can approximate the anyon as an object occupying the two sites and the nearest neighbors of either of them (see the dashed red circles in Fig.\ \ref{fig:braiding_path1} (d)).

To check how good this approximation is, we introduce the local excess particle density 
\begin{equation}
\rho_{\mathrm{NN}1}(t)=\sum_{j}\theta_{j,1}(T) \rho_j
\label{eq:rhoNN}
\end{equation}
where 
\begin{multline}
    \theta_{j,1}(T) =\\ \left\{\begin{array}{ll}
        1, & \text{if } |z_j-z_{k(T)}|\leq 1~\mathrm{or}~|z_j-z_{k(T+1)}|\leq 1\\
        0, & \text{otherwise } \\
        \end{array}\right..
\end{multline}
Analogously, we define $\rho_{\mathrm{NN}2}(t)$ and $\theta_{i,2}(T)$ by focusing on the second anyon, i.e. replacing $k(T)$ by $l(T)$. In the situation from Fig. \ref{fig:braiding_path1} (b), the anyons can be separated only when $\rho_{\mathrm{NN}1}(t)$ is close to (or ideally, equal to) $-0.5$. Note that when $k(T)\neq k (T+1)$, \eqref{eq:rhoNN} counts the excess particle density around both sites $k(T)$ and $k (T+1)$, even when $s=0$ and the anyon is centered at site $k(T)$. But this does not change the argument that in order to separate the anyons, we should have $\rho_{\mathrm{NN}1}(t)\approx -0.5$.

The results are shown in Fig.\ \ref{fig:braiding_path1} (e) (the exchange path) and Fig. \ref{fig:braiding_path1} (f) (the AB path) with red markers and lines. The dashed (solid) lines correspond to $\rho_{\mathrm{NN}1}(t)$ ($\rho_{\mathrm{NN}2}(t)$). The four sites on which the anyons are closest to each other in the AB phase calculation are denoted by gray lines in Fig. \ref{fig:braiding_path1} (f). It can be seen that $\rho_{\mathrm{NN}1}(t)$ and $\rho_{\mathrm{NN}2}(t)$ can depart quite far from the perfect value $-0.5$. Also, the result seems to depend on the position of the anyon on the path (although for the static anyon in the AB phase calculation, $\rho_{\mathrm{NN}2}(t)$ seems quite stable). Therefore, the assumption that the anyon is an object occupying only two sites and their nearest neighbors is a relatively rough approximation.  Hence, we should not expect that the braiding phase would be perfectly equal to $\pi/2$.

\begin{figure}[h]
\includegraphics[width=0.45\textwidth]{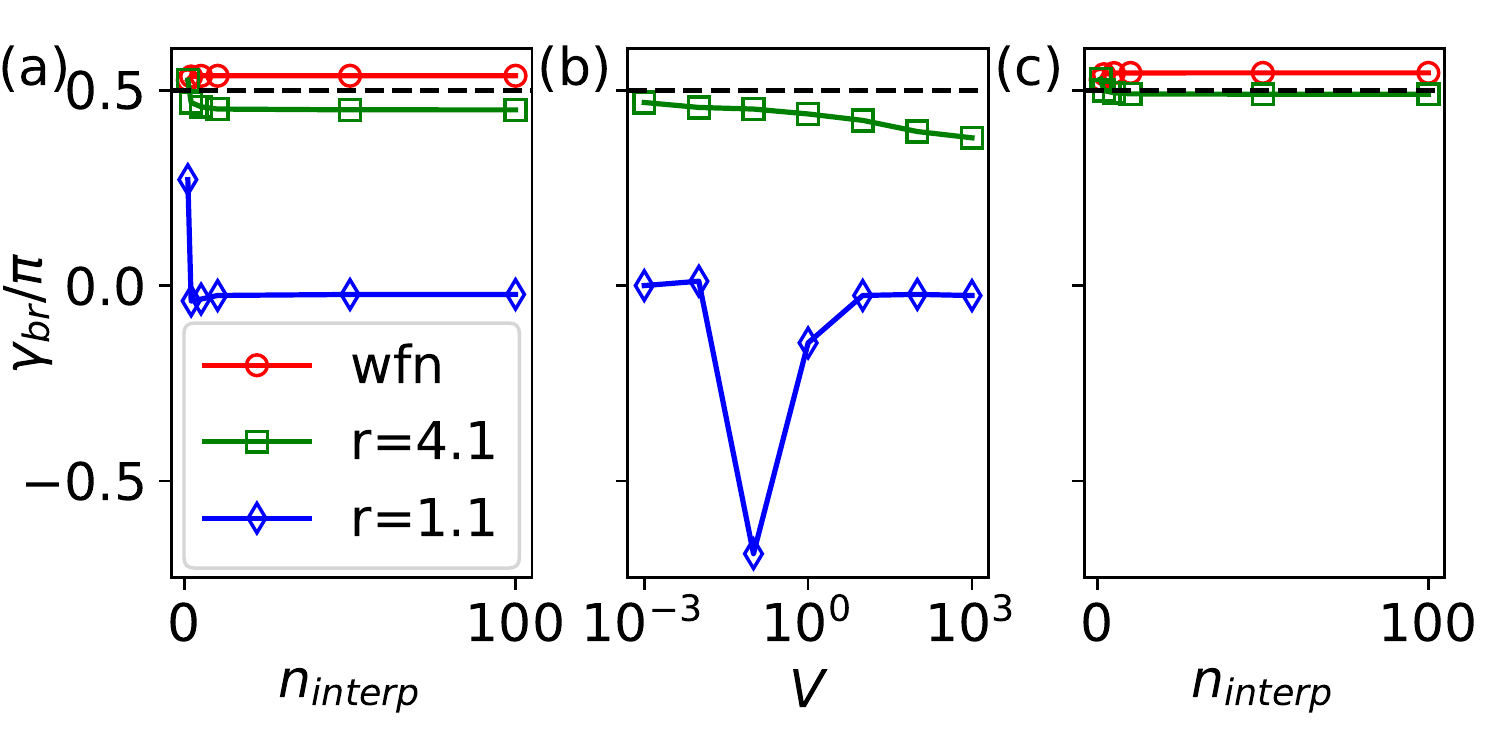}
\caption{Braiding phases for the triangle with $M_0=6$ and $M=5$. (a) and (b) show the results for the path depicted in Fig.\ \ref{fig:braiding_path1}, while (c) refers to the path from Fig.\ \ref{fig:braiding_path2}. In (a) and (c), the phase is shown as a function of the number $n_{\mathrm{interp}}$ of interpolating steps between sites, and in (b) as a function of the potential strength $V$ at constant $n_{\mathrm{interp}}=10$. See the main text for further details.}
\label{fig:braiding_phases}
\end{figure}

The phase $\gamma_{\mathrm{br}}$ as a function of the number of interpolation steps $n_{\mathrm{interp}}$ is shown in Fig.\ \ref{fig:braiding_phases} (a) using red markers and lines. For $n_{\mathrm{interp}}=100$, we obtain a result $\gamma_{\mathrm{br}}=0.537\pi$, which is close to $\pi/2$ (black dashed line) but still there is a notable discrepancy. We cannot separate the anyons further and check whether the discrepancy decreases. Thus, while the result of the braiding operation in this system can hint at the presence of anyons, the system is too small to unambiguously demonstrate it, at least as long as the ground state is described by model wavefunctions \eqref{eq:anyons}. The full excess particle density distribution at any point of either path can be seen in the animations provided in the Supplementary Material \cite{supplementary2}.

We do not expect that the results in the systems described by a Hamiltonian will be clearer. Nevertheless, we can ask: how well do they reproduce the results for a model wavefunction? We again use the paths from Figs.\ \ref{fig:braiding_path1} (a) and (b). For the $s=0$ cases, we use two potentials with strength $V$ at sites $k(T)$, $l(T)$. To interpolate between the sites, we use the following scheme,
\begin{multline}
H'(t)=H+\left(1-\lambda \left( s \right) \right) V n_{k(T)}+ \lambda(s) V n_{k(T+1)} +  \\
\left(1-\lambda \left( s \right) \right) V n_{l(T)}+ \lambda(s) V n_{l(T+1)},
\label{eq:potential_interp}
\end{multline}
where $\lambda(s)=s-\sin(2\pi s)/(2\pi)$. Note that because either $k(T)=k(T+1)$ or $l(T)=l(T+1)$, one of the potentials remains static at each point of the path.

We first consider the nonlocal case $r=4.1$, for which we set $V=0.1$. This potential is much smaller than the $V=1000$ used in Sec.\ \ref{sec:anyons}. While an arbitrarily high potential can be used to pin the anyons to sites (the higher the better: the model wavefunction has $\braket{n_j}=0$ if $w_k=z_j$), high potentials raise problems for interpolation. At a sufficiently high $V$, applying \eqref{eq:potential_interp} to interpolate between sites $k(T)$ and $k(T+1)$ would lead to high potentials on both of these sites for every $s\neq 0$, enforcing both $\braket{n_{k(T)}}\approx 0$ and $\braket{n_{k(T+1)}}\approx 0$. Then, there would be a significant difference in the wavefunction at $s=0$ and $s=1$ but not much difference between $s=1$ and other $s\neq 0$. Therefore, to make the interpolation more smooth, we choose a much smaller $V$ than in Sec.\ \ref{sec:anyons}.

In Fig.\ \ref{fig:braiding_path1} (e) and (f), as well as in Fig.\ \ref{fig:braiding_phases} (a) and (b), these results are plotted using green lines and markers. When each anyon is pinned to one site (i.e. $s=0$), the results for the $r=4.1$ Hamiltonian and the model wavefunction are almost the same -- see the red and green curves intersecting at integer $t$ in Fig.\ \ref{fig:braiding_path1} (e) and (f), as well as the red and green markers coinciding at $n_{\mathrm{interp}}=1$ in Fig.\ \ref{fig:braiding_phases} (a). However, because the methods of interpolation between the sites are different in the two cases, the curves in Fig.\ \ref{fig:braiding_path1} (e) and (f) depart from each other at noninteger $t$. The Hamiltonian case displays larger excess particle density variations than the model wavefunction case. In particular, in the AB phase calculation, $\rho_{\mathrm{NN}1}(t)$ reaches almost $-0.7$ at some point, which is far from the ideal $-0.5$ value (see also the animations in the Supplementary Material \cite{supplementary2}). The braiding phase shown in Fig.\ \ref{fig:braiding_phases} (a) also differs between the two cases. We have $\gamma_{\mathrm{br}}=0.450 \pi$ at $n_{\mathrm{interp}}=100$ for the Hamiltonian case, which is further away from $0.5\pi$ than the result for the model wavefunction. Nevertheless, this is still relatively close to $\pi/2$. We also plot $\gamma_{\mathrm{br}}$ for $n_{\mathrm{interp}}=10$ and different values of $V$ ranging from $V=0.001$ to $V=1000$ in Fig.\ \ref{fig:braiding_phases} (b), showing that the deviation from the ideal value increases with the strength of the potential, but the phase remains relatively close to $\pi/2$ for a range of $V$ values on the left side of the plot.

For comparison, we also repeat the calculations for the $r=1.1$ Hamiltonian with $V=10$. These results are shown in Fig.\ \ref{fig:braiding_path1} (e) and (f), as well as in  Fig.\ \ref{fig:braiding_phases} (a), using blue markers and lines. In Fig.\ \ref{fig:braiding_path1} (e) and (f) we can see that $\rho_{\mathrm{NN}1}(t)$ and $\rho_{\mathrm{NN}2}(t)$ vary very strongly. In the animations shown in the Supplementary Material \cite{supplementary2}, one can see that the excess particle density patterns change abruptly. The braiding phase plotted in Fig.\ \ref{fig:braiding_phases} (a) is close to 0 for high enough $n_{\mathrm{interp}}$, showing no indication of fractional statistics. This may be connected to quite small overlaps with model states (in general smaller than the one in Fig.\ \ref{fig:OverlapQuasiholes}, e.g. for one of the steps we have $|\braket{\psi|\tilde{\Psi}}|^2\approx0.492$), but it is also possible that the correct anyonic statistics would be visible in larger $r=1.1$ systems even if the excitations are not described by \eqref{eq:anyons}. We performed the computations also for various other values of $V$, keeping $n_{\mathrm{interp}}=10$. The result, shown in Fig.\ \ref{fig:braiding_phases} (c), depends on $V$, and is not close to $\pi/2$ for any of the considered cases.  

\begin{figure}[h]
\includegraphics[width=0.45\textwidth]{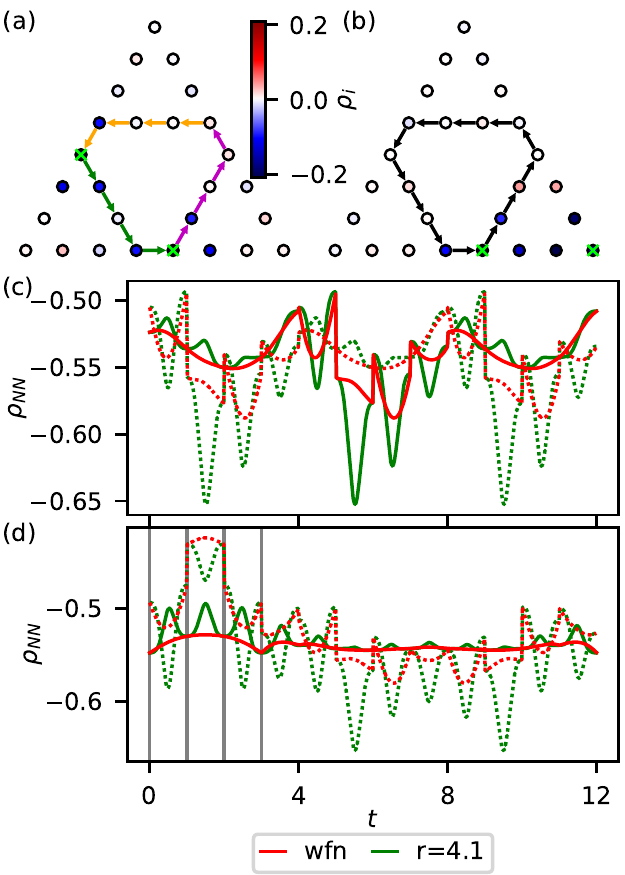}
\caption{The second variant of a braiding path in a triangle with $M_0=6$, $M=5$. (a), (b): the excess particle density \eqref{eq:excesscharge} of the model wavefunction at the beginning of the exchange and braiding path, respectively. The bright green crosses denote the anyon positions $w_k$. The paths are denoted by arrows.  (c), (d) The local excess particle throughout the exchange and AB paths, respectively. The dashed and solid lines correspond to $\rho_{\mathrm{NN}1}(t)$ and $\rho_{\mathrm{NN}2}(t)$, respectively, while different colors denote different cases (the model wavefunction and the $r=4.1$ Hamiltonian).  See the main text for more details.}
\label{fig:braiding_path2}
\end{figure}

In addition, we also study a different path, shown in Fig.\ \ref{fig:braiding_path2} (a), (b). For the exchange path, the anyons are again moved along the arrows in the following order: purple, green, orange. The braiding phases are shown in Fig.\ \ref{fig:braiding_phases} (b). For $n_{\mathrm{interp}}=100$, we obtain $\gamma_{\mathrm{br}}=0.544\pi$ for the model wavefunction and $\gamma_{\mathrm{br}}=-0.489 \pi$ for an $r=4.1$ Hamiltonian. The similarity between the braiding phases in the two paths suggests the statistical origin of the phase. However, as seen in Fig.\ \ref{fig:braiding_path2} (c) and (d), $\rho_{\mathrm{NN}1}(t)$ and $\rho_{\mathrm{NN}2}(t)$ again vary throughout the path and depart quite strongly from the ideal value $-0.5$  (for example, both exceed $-0.65$ at some $t_j$ values on the exchange path for $r=4.1$ Hamiltonian). This again showcases the problems with separating the anyons and, in consequence, with evaluating the statistics.

In summary, we have shown that our systems are too small to convincingly demonstrate fractional statistics of anyons. Even in the case of the model wavefunction, the finite-size effects are notable, because the anyons cannot be separated far enough from each other. For a system described by a nonlocal Hamiltonian, the local density depletions (which we expect to be anyons due to the similarity with the model state) are even harder to separate, although the results display some similarity to the results from the model wavefunction. For a local Hamiltonian, the behavior of the system within the braiding process bears no resemblance to the behavior of the model wavefunction. Nevertheless, the braiding phases roughly close to $\pi/2$ arising on two different paths in the case of the nonlocal Hamiltonian suggest that their origin might be statistical.

We note that on the carpet (in the cases we can study with exact diagonalization) the possibilities of separating anyons are even worse. Due to smaller $\eta$, the quasiholes are considerably bigger compared to the distance between nearest-neighboring sites, as one can see in Fig.\ \ref{fig:DensityDifference} (c) (where the excess charge density is quite similar to the one for the model wavefunction). At the same time, due to the lattice structure, the distance at which they can be separated in the most problematic point of the AB path is only slightly larger.

\section{Conclusions}\label{sec:conclusions}

We have numerically constructed parent Hamiltonians for Laughlin states in fractal lattices. All the Hamiltonians have the form of a tight-binding model of hardcore bosons, resembling a Hofstadter or Kapit-Mueller model. It is possible to get reasonable overlaps ($|\braket{\psi|\Psi}|^2>0.81$) between the ground state and the model wavefunction even in local models with up to third-neighbor hopping (and, for small enough number of particles, also for models with even smaller maximum hopping distance). For up to $M_0=4$, we also obtain overlaps $|\braket{\psi|\tilde{\Psi}}|^2>0.9$ between the ground state with additional potentials and the model wavefunction with two quasiholes. In the case of nonlocal, Kapit-Mueller-like models, the overlap with a model wavefunction can be nearly perfect, both for the case with no anyons and with quasiholes. 

Moreover, we investigated the braiding process for an $M_0=6$ triangle. We analyzed the finite-size effects in the model wavefunction (which displays fractional statistics clearly in large fractal lattices \cite{manna2020anyon}), showing they distort the braiding phase in the small system considered in this work and do not allow to separate the anyons clearly. Then, we compared the braiding process for a model wavefunction and a nonlocal Hamiltonian, showing some similarity between these cases (including braiding phases relatively close to $\pi/2$ on two paths in both cases), providing a reason for hope that a similar Hamiltonian might allow to observe fractional statistics unambiguously for a larger structure.

We intend the results obtained in this work to be a bridge between the model wavefunctions and tight-binding models. Our Hamiltonians are considerably simpler than the exact parent Hamiltonians proposed before (even in the nonlocal case), while still retaining the connection with model wavefunctions.

We note that within the approach used in this work the Hamiltonians have to be determined separately for each lattice and system size. Thus, similarly to the exact Hamiltonian from \cite{manna2020anyon}, they depend on the size and shape of the system. In this work, we concentrated on system sizes available in exact diagonalization. However, one can use the method also for bigger systems. The procedure of finding a parent Hamiltonian requires operating on the full many-body basis of the system, composed of $\binom{N}{M}$ states, but does not require diagonalizing any matrix of size $\binom{N}{M}\times\binom{N}{M}$. Thus, while the system size is limited, the limitation is less strict than in the exact diagonalization procedure.

We expect that the method used in our paper can be applied to the non-Abelian bosonic Moore-Read states as well. Bosonic lattice Moore-Read states were defined for fractal lattices with 3-particle onsite hardcore interaction (sites occupied with up to two particles) \cite{manna2020quasiparticles}. Analogy with the Kapit-Mueller model in two dimensions \cite{kapit2012nonabelian} suggests that combining such an interaction with single-particle terms may be enough to construct a parent Hamiltonian for this state.

\acknowledgments

We thank Callum W. Duncan for discussions. This work has been supported by the Independent Research Fund Denmark under grant number 8049-00074B and the Carlsberg Foundation under grant number CF20-0658.

\bibliography{fractal}

\end{document}